\documentclass[12pt,preprint]{aastex}

\shorttitle{The Spatial Distribution of COMs in L1544}
\shortauthors{Jim\'enez-Serra et al.}

\begin{document}


\title{The Spatial Distribution of Complex Organic Molecules in the L1544 Pre-stellar Core}


\author{Izaskun Jim\'{e}nez-Serra\altaffilmark{1,2,3}, Anton I. Vasyunin\altaffilmark{4,5}, Paola Caselli\altaffilmark{4}, Nuria Marcelino\altaffilmark{6,7}, Nicolas Billot\altaffilmark{8}, Serena Viti\altaffilmark{2}, Leonardo Testi\altaffilmark{3,9}, Charlotte Vastel\altaffilmark{10,11}, Bertrand Lefloch\altaffilmark{12,13} and Rafael Bachiller\altaffilmark{14}}

\altaffiltext{1}{Astronomy Unit, School of Physics \& Astronomy, Queen Mary University of London,
Mile End Road, E1 4NS, London (UK); i.jimenez-serra@qmul.ac.uk}
\altaffiltext{2}{Department of Physics \& Astronomy, University College London, 132 Hampstead Road, NW1 2PS London (UK)}
\altaffiltext{3}{European Southern Observatory (ESO), Karl-Schwarzschild-Str. 2, 85748 Garching (Germany)}
\altaffiltext{4}{Max-Planck-Institut f\"ur extraterrestrische Physik (MPE), Gie{\ss}enbachstr., 85741 Garching (Germany)}
\altaffiltext{5}{Ural Federal University, Ekaterinburg, Russia}
\altaffiltext{6}{INAF, Osservatorio de Radioastronomia, Via P. Gobetti 101, 40129 Bologna, (Italy)}
\altaffiltext{7}{Instituto de Ciencia de Materiales de Madrid, CSIC, C/ Sor Juana In\'es de la Cruz 3, 28049 Cantoblanco (Spain)}
\altaffiltext{8}{Instituto de Radioastronom\'{\i}a Milim\'etrica, Avenida Divina Pastora 7, E-18012 Granada (Spain)}
\altaffiltext{9}{INAF-Osservatorio Astrofisico di Arcetri, Largo E. Fermi 5, I-50125 Firenze, Italy}
\altaffiltext{10}{Universit\'e de Toulouse, UPS-OMP, IRAP, F-31400 Toulouse, France}
\altaffiltext{11}{CNRS, IRAP, 9 Av. Colonel Roche, BP 44346, F-31028 Toulouse Cedex 4, France}
\altaffiltext{12}{Univ. Grenoble Alpes, IPAG, F-38000 Grenoble, France}
\altaffiltext{13}{CNRS, IPAG, F-38000 Grenoble, France}
\altaffiltext{14}{Observatorio Astron\'omico Nacional (OAN, IGN), Calle Alfonso XII 3, E-28014 Madrid, Spain}

\begin{abstract}

The detection of complex organic molecules (COMs) toward cold sources such as pre-stellar cores (with T$<$10 K), has challenged our understanding of the formation processes of COMs in the interstellar medium. Recent modelling on COM chemistry at low temperatures has provided new insight into these processes predicting that COM formation depends strongly on parameters such as visual extinction and the level of CO freeze out. We report deep observations of COMs toward two positions in the L1544 pre-stellar core: the dense, highly-extinguished continuum peak with A$_V$$\geq$30$\,$mag within the inner 2700$\,$au; and a low-density shell with average A$_V$$\sim$7.5-8$\,$mag located at 4000$\,$au from the core's center and bright in CH$_3$OH. Our observations show that CH$_3$O, CH$_3$OCH$_3$ and CH$_3$CHO are more abundant (by factors $\sim$2-10) toward the low-density shell than toward the continuum peak. Other COMs such as CH$_3$OCHO, c-C$_3$H$_2$O, HCCCHO, CH$_2$CHCN and HCCNC show slight enhancements (by factors $\leq$3) but the associated uncertainties are large. This suggests that COMs are actively formed and already present in the low-density shells of pre-stellar cores. The modelling of the chemistry of O-bearing COMs in L1544 indicates that these species are enhanced in this shell because i) CO starts freezing out onto dust grains driving an active surface chemistry; ii) the visual extinction is sufficiently high to prevent the UV photo-dissociation of COMs by the external interstellar radiation field; and iii) the density is still moderate to prevent severe depletion of COMs onto grains.
\end{abstract}


\section{Introduction}

Complex Organic Molecules (COMs) are carbon-based species with $\geq$6 atoms in their molecular structure \citep[][]{her09}. The most prolific regions in the detection of COMs in the interstellar medium (ISM) have been massive hot cores and Giant Molecular Clouds in the Galactic Center \citep[SgrB2 (N) and (M);][]{holl00,holl06,req08,bell08,bell14} and low-mass hot corinos \citep[IRAS16293-2422;][]{cec00,bott04,jor12}. Until recently, it was believed that COMs form on dust grains via hydrogenation \citep[][]{char95} or radical-radical reactions favoured by the heating from the central protostar \citep[at T$\geq$30$\,$K; see][]{garr08}. However, the detection of COMs such as propylene (CH$_2$CHCH$_3$), acetaldehyde (CH$_3$CHO), dimethyl ether (CH$_3$OCH$_3$) or methyl formate (CH$_3$OCHO) in dark cloud cores and pre-stellar cores with T$\leq$10$\,$K (B1-b, TMC-1, L1689B or L1544) has recently challenged our understanding of COM formation \citep[][]{mar07,oberg10,bac12,cer12,vas14,loi16}. 

Several mechanisms have been proposed to explain the presence of COMs in cold cores: gas-phase formation, non-canonical chemical explosions, cosmic-ray induced radical diffusion, impulsive spot heating of grains, or radical-radical recombination after H-atom addition/abstraction reactions on grain surfaces \citep[][]{vasy13,raw13,bal15,reb14,ivlev15,chu16}. However, information about the spatial distribution of COMs in cold cores is lacking (as well as of their radial abundance profile probing different density and extinction regimes), which prevents us from testing these COM formation scenarios. 

The detection of a low-density, CH$_3$OH-rich shell around the continuum peak of the L1544 pre-stellar core \citep[][]{biz14} offers a unique opportunity to test COM formation scenarios in cold sources. Species such as C$_3$O, ketene (H$_2$CCO), formic acid (HCOOH) and acetaldehyde may be spatially co-located to CH$_3$OH in L1544 and may form at this low-density shell \citep{vas14}. We report high-sensitivity observations of COMs toward two positions in the L1544 pre-stellar core: the continuum peak and a position within the low-density, CH$_3$OH-rich shell reported by \citet[][hereafter the {\it CH$_3$OH peak}]{biz14}. Our results suggest that COMs are actively formed in the low-density shells of pre-stellar cores\footnote{Part of these observations belong to the ASAI (Astrochemical Surveys at IRAM) program.}.

\section{Observations}
\label{obs}

High-sensitivity single-pointing 3$\,$mm observations were carried out toward two positions in the L1544 core during 10-16 December 2014 and 15-16 June 2016 with the Instituto de Radioastronom\'{\i}a Milim\'etrica (IRAM) 30$\,$m telescope. We pointed our observations toward the core's dust continuum peak, $\alpha$(J2000)=5$^h$04$^m$17.21$^s$, $\delta$(J2000)=25$^\circ$10$'$42.8$"$, and toward the {\it CH$_3$OH peak}, $\alpha$(J2000)=5$^h$04$^m$18$^s$ and $\delta$(J2000)=25$^\circ$11$'$10$"$ \citep[][]{biz14}. This position is $\sim$30$"$ away from the core's center \citep[$\sim$0.02$\,$pc or 4000$\,$au at 140$\,$pc;][]{elias78}. 

The observations were done in wobbler-switched mode with an angular throw of $\pm$120$"$. The dual sideband (2SB) EMIR E090 receivers were tuned at 84.37$\,$GHz and 94.82$\,$GHz with rejections $\geq$10$\,$dB. We used three slightly different central frequencies (84.36, 84.37, 84.38$\,$GHz and 94.80, 94.82, 94.84$\,$GHz) to avoid the appearance of weak spurious features in the spectra. The FTS spectrometer allowed us to observe the inner part of each 4$\,$GHz sub-bands covering 7.2$\,$GHz in total, and it provided a spectral resolution of 50$\,$kHz (0.15-0.18$\,$km s$^{-1}$ at 3$\,$mm). The observed frequency ranges were 83.4-85.2, 86.7-88.5, 99.1-100.9, 102.4-104.2$\,$GHz and 78.2-80.0, 81.4-83.2, 93.9-95.7, 97.1-98.9$\,$GHz. Typical system temperatures were 70-120$\,$K. The beam size was $\sim$28-31$"$ between 79-87$\,$GHz, and $\sim$24-26$"$ between 94-103$\,$GHz. Intensities were calibrated in units of antenna temperature, T$_{\rm A}^*$, and converted into main beam temperature, T$_{\rm mb}$, by using beam efficiencies of 0.81 at 79-100$\,$GHz, and 0.78 at 103$\,$GHz. The RMS noise level was 1.6-2.8 mK for the core's center position, and 2.2-3.7 mK for the {\it CH$_3$OH peak}.  

\section{Results}
\label{res}

\begin{figure}
\begin{center}
\includegraphics[angle=270,width=0.9\textwidth]{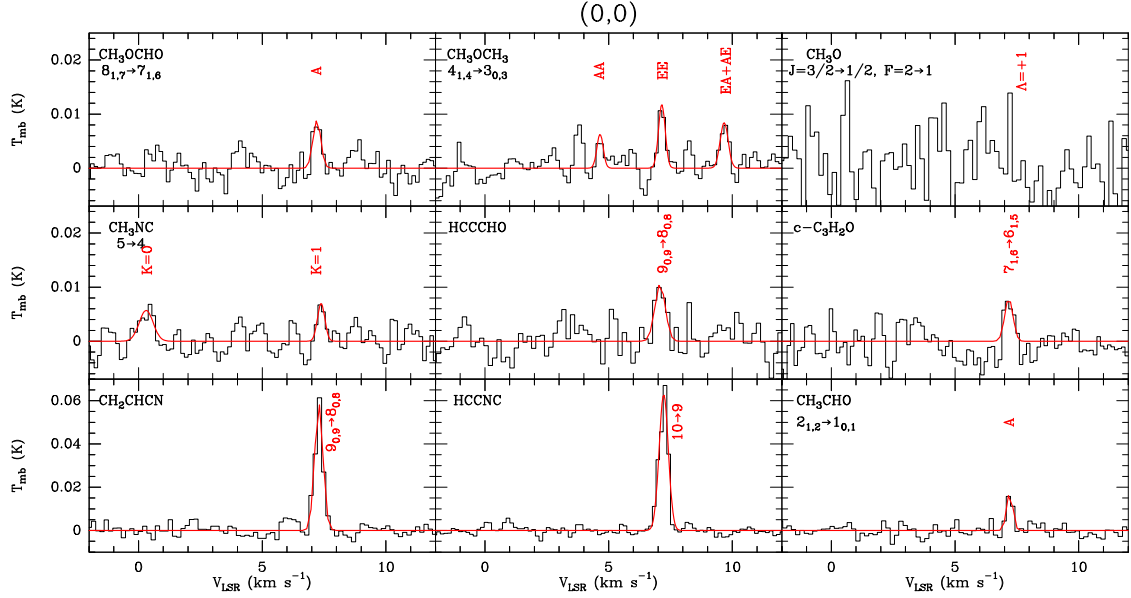}
\caption{Sample of COM lines detected toward L1544's dust continuum peak. Red lines show the Gaussian line fits derived for the COM transitions (see Table$\,$\ref{tab2}). The CH$_3$O spectrum has been extracted from the ASAI data \citep{vas14}.}
\label{fig1}
\end{center}
\end{figure}

\begin{figure}
\begin{center}
\includegraphics[angle=270,width=0.9\textwidth]{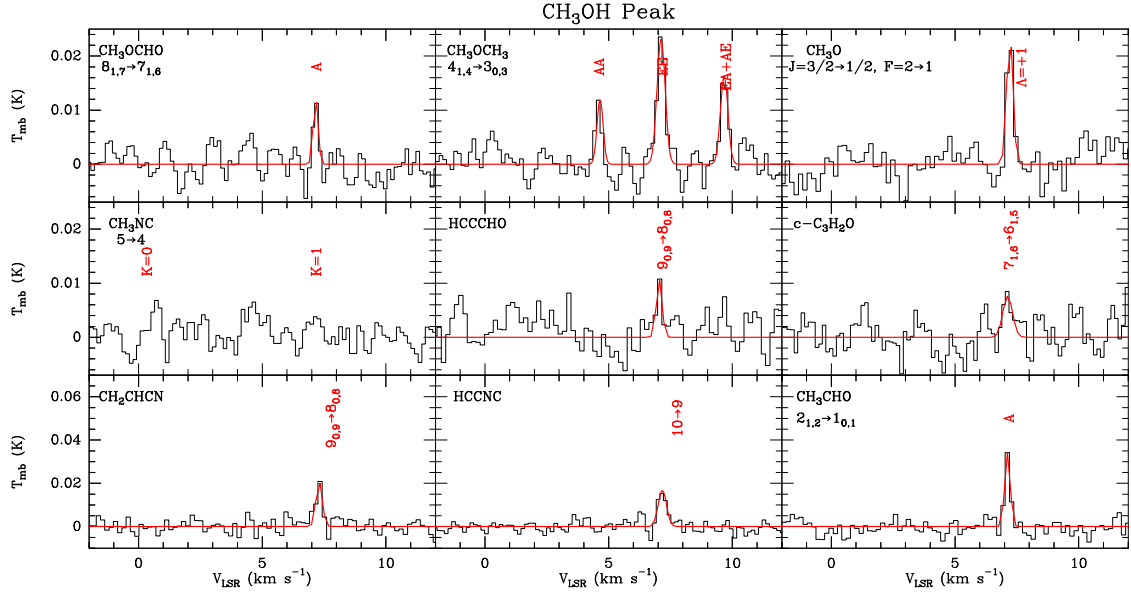}
\caption{As in Figure$\,$\ref{fig1}, but for the position of the {\it CH$_3$OH peak}.} 
\label{fig2}
\end{center}
\end{figure}

Figures$\,$\ref{fig1} and \ref{fig2} show some of the COM transitions detected toward L1544, while Table$\,$\ref{tab2} lists all observed transitions and their derived line parameters. For completeness, we also provide the information about the covered transitions of acetaldehyde (CH$_3$CHO) -- species already reported by \citet{vas14} -- and the upper limits of formamide (NH$_2$CHO), methyl isocyanate (CH$_3$NCO) and glycine (NH$_2$CH$_2$COOH). For all COMs the spectroscopic data were extracted from the  JPL catalog \citep[][]{pick98}, except for dimethyl ether (CH$_3$OCH$_3$) for which we used SLAIM\footnote{The Spectral Line Atlas of Interstellar Molecules is available at http://www.splatalogue.net \citep[][]{rem07}.}, and cyclopropenone (c-C$_3$H$_2$O), methyl isocyanate and glycine, for which we used the CDMS catalog \citep[][]{mull05}. 

Our high-sensitivity spectra reveal the detection of large COMs such as methyl formate (CH$_3$OCHO) and dimethyl ether (CH$_3$OCH$_3$) that had remained elusive in previous campaigns \citep[see upper limits in][]{vas14}. Other species detected are propynal (HCCCHO), cyclopropenone (c-C$_3$H$_2$O), ethynyl isocyanide (HCCNC) and vinyl cyanide (CH$_2$CHCN; Figures$\,$\ref{fig1} and \ref{fig2}). Methoxy (CH$_3$O), considered as a COM precursor or COM dissociation product, is also detected toward the {\it CH$_3$OH peak}. All these molecules are observed at the $\sim$10-20$\,$mK intensity level (Table$\,$\ref{tab2}). In Figure$\,$\ref{fig1}, we also report the tentative detection of methyl isocyanide (CH$_3$NC) toward the continuum peak. Its K=0,1 lines are detected at the 2.5$\sigma$ and 3.2$\sigma$ level. We are confident about the correct identification of all transitions since their derived radial velocities match the $\rm v_{\rm LSR}$ of the source (7.2$\,$km$\,$s$^{-1}$), and for most species at least two transitions are detected. In addition, we have looked for other lines that could be blended and have found none. Since the COM line profiles are narrow (linewidths $\sim$0.3-0.4$\,$km$\,$s$^{-1}$; Table$\,$\ref{tab2}), it is unlikely that they appear blended with unknown species.

Table$\,$\ref{tab2} shows that the line emission from O-bearing species is either brighter toward the {\it CH$_3$OH peak} (by factors of $\sim$2-3 for CH$_3$O, CH$_3$CHO, CH$_3$OCHO and CH$_3$OCH$_3$), or remains constant toward both positions (HCCCHO and c-C$_3$H$_2$O). The N-bearing COMs HCCNC, CH$_3$NC and CH$_2$CHCN, on the contrary, show an opposite behaviour with brighter emission seen toward the center of the core. As shown in Section$\,$\ref{abun}, this behaviour translates into larger enhancements toward the {\it CH$_3$OH peak} for CH$_3$O, CH$_3$CHO and CH$_3$OCH$_3$ than for the rest of COM species. Formamide, methyl isocyanate and glycine are not detected toward any position in L1544 (Table$\,$\ref{tab2}). 

\section{Excitation analysis of COMs}
\label{exc}

For CH$_3$O, CH$_3$OCHO, CH$_3$OCH$_3$, CH$_3$CHO, CH$_3$NC and CH$_2$CHCN, we have detected several lines so that a multi-line excitation analysis can be performed. The excitation temperature, T$_{\rm ex}$, and total column density, N$_{obs}$, of these molecules have been calculated using the MADCUBAIJ software \citep[][]{smar11,riv16}, assuming extended emission and LTE conditions. Except for CH$_3$NC (which is tentatively detected; Section $\,$\ref{res}), the derived T$_{\rm ex}$ is $\sim$5-6$\,$K toward the core's center and $\sim$5-8$\,$K toward the {\it CH$_3$OH peak} (Table$\,$\ref{tab3}). Errors in Table$\,$\ref{tab3} correspond to 1$\sigma$ uncertainties. 
We note that in some cases we had to fix T$_{\rm ex}$ to make MADCUBAIJ converge and find a solution. Since c-C$_3$H$_2$O is not included in MADCUBAIJ, we used the Weeds software for this molecule \citep{maret11} and assumed that its emission fills the beam.  

For the COMs with only one transition detected (and also for the non-detections), we again used MADCUBAIJ and assumed extended emission and a T$_{\rm ex}$ range of 5-10$\,$K for both positions. This T$_{\rm ex}$ range agrees well with the values of T$_{\rm ex}$ obtained from the COM multi-line excitation analysis (Table$\,$\ref{tab3}), and with the T$_{\rm ex}$ measured from CH$_3$OH by \citet[][]{biz14}. The estimated values of N$_{obs}$ are reported in Table$\,$\ref{tab3}.

\section{COM abundance profiles in L1544}
\label{abun}

By calculating the COM molecular abundances toward the two positions in L1544, we can provide constraints to the COM abundance profiles as a function of radius within the core. For the continuum peak, we need to calculate the H$_2$ column density, N(H$_2$), within a radius of 13$"$ or 1900$\,$au (i.e. half the IRAM 30$\,$m beam of our observations). The N(H$_2$) of the core for radii $\leq$2500$\,$au ($\sim$18$"$ at 140$\,$pc) is nearly flat with a radius dependence N(H$_2$)$\propto r^{-0.8}$ \citep[][]{ward99}. By using the peak N(H$_2$) obtained by \citet[][9.4$\times$10$^{22}$$\,$cm$^{-2}$]{cra05} for a radius of 6.5$"$ ($\sim$910$\,$au), we derive that N(H$_2$)=5.4$\times$10$^{22}$$\,$cm$^{-2}$ for a radius of 13$"$ (1900$\,$au). This value is similar to that estimated by \citet{bac00} for the same radius (4.5$\times$10$^{22}$$\,$cm$^{-2}$; see their Table$\,$2). The slight difference is due to the dust temperatures assumed in both calculations (12.5$\,$K in Bacmann et al. 2000 vs. 10$\,$K in Crapsi et al. 2005). Hereafter, we use N(H$_2$)=5.4$\times$10$^{22}$$\,$cm$^{-2}$ for the position of the continuum peak within a radius of 13$"$.

For the {\it CH$_3$OH peak}, we assume an H$_2$ column density of 1.5$\times$10$^{22}$$\,$cm$^{-2}$ as derived by \citet{spez16} from Herschel data. The latter N(H$_2$) corresponds to a visual extinction $A_V$$\sim$15-16$\,$mag using the \citet{boh78} formula. We note, however, that the model of L1544 by \citet{ket10} and \citet{ket14} considers the extinction as a function of radius (within the core and not along the line-of-sight) and therefore the modelled $A_V$ at this position is about half that measured along the line-of-sight ($A_V$$\sim$7.5-8$\,$mag; see also Section$\,$\ref{model}).

From Table$\,$\ref{tab3} and Figure$\,$\ref{fig3} (lower panel), we find that CH$_3$O, CH$_3$OCH$_3$, CH$_3$CHO are enhanced respectively by factors $\geq$4-5, $\sim$2 and $\sim$10 in the {\it CH$_3$OH peak} with respect to L1544's center (note that their associated 1$\sigma$ uncertainties are lower than these enhancements). The other O-bearing COMs CH$_3$OCHO, c-C$_3$H$_2$O and HCCCHO show average abundances slightly higher toward the {\it CH$_3$OH peak} than toward the core center (by factors $\leq$3), although they agree within the uncertainties. The same applies to the N-bearing COMs CH$_2$CHCN, CH$_3$NC and HCCNC, whose abundances lie within the uncertainties (Figure$\,$\ref{fig3}). 
 

\section{Upper limits to the abundance of pre-biotic COMs}
\label{peptide}

The high-sensitivity spectra obtained toward L1544 allow us to provide stringent upper limits to the abundance of pre-biotic COMs such as glycine, NH$_2$CHO and CH$_3$NCO. As shown in Table$\,$\ref{tab3}, the derived upper limits to the column density of glycine are factors 40-120 lower than the best upper limits obtained toward the Galactic Center \citep[$\leq$4$\times$10$^{14}$$\,$cm$^{-2}$;][]{jon07}. Our most stringent upper limit to the abundance of glycine in L1544 ($\leq$6$\times$10$^{-11}$; Table$\,$\ref{tab3}) is also a factor of 5 lower than that inferred for the outer envelope of IRAS16293-2422 \citep[$\leq$3$\times$10$^{-10}$;][]{cec00}. Stacking analysis of the glycine lines with similar expected intensities in our frequency setup would reduce the rms noise level by a factor of $\sim$3 (10 lines are covered), which implies an upper limit of $\leq$2$\times$10$^{-11}$. This upper limit is close to the glycine abundance assumed by \citet[][]{jim14} for the detectability of this molecule in pre-stellar cores ($\sim$3$\times$10$^{-11}$). The upper limits derived for NH$_2$CHO [$\leq$(2.4-8.7)$\times$10$^{-13}$] and CH$_3$NCO [$\leq$(0.2-4.2)$\times$10$^{-11}$; Table$\,$\ref{tab3}] are consistent with those measured in L1544 and B1-b by \citet[][$\leq$5$\times$10$^{-13}$]{lop15} and \citet[][$\leq$2$\times$10$^{-12}$]{cer16} respectively.

\section{Chemical modelling of O-bearing COMs}
\label{model}

In Section$\,$\ref{abun}, we have reported abundance enhancements by factors $\sim$2-10 for CH$_3$O, CH$_3$OCH$_3$ and CH$_3$CHO as a function of distance in L1544. Other COMs such as CH$_3$OCHO may also be enhanced at larger radii although its derived abundances agree within the uncertainties (Section$\,$\ref{abun}). In this Section, we model the chemistry of O-bearing COMs\footnote{Except NH$_2$CHO, none of the N-bearing COMs reported here are currently included in the chemical network of Vasyunin et al. (in prep.).} in L1544 by using the 1D physical structure derived by \citet{ket10} and \citet{ket14}. 

We use the chemical code of \citet{vasy13}, which considers that COMs are formed via gas-phase ion-neutral and neutral-neutral reactions after the release of precursor molecules from dust grains via chemical reactive desorption. This code has been updated with a new multilayer treatment of ices, an advanced treatment of reactive desorption based on the experiments by \citet{min16}, and new gas-phase reactions proposed by \citet{sha13} and \citet{bal15}. The complete results from this modelling will be reported in Vasyunin et al., (in prep.). 

The chemical evolution of COMs is followed over 3$\times$10$^6$$\,$years toward 129 points along the radius of the core to a distance $\sim$65500$\,$au. For all these positions, the density, temperature and A$_v$ are taken from \citet{ket14}. The initial molecular abundances are calculated by simulating the chemistry of a translucent cloud with density n(H)=10$^2$ cm$^{-3}$ and T$_{gas}$=T$_{dust}$=20 K over 10$^6$$\,$years. The initial atomic abundances are the same as those of \citet{vasy13}.

Our model shows that the abundances of COMs such as CH$_3$OCHO and CH$_3$OCH$_3$ change dramatically with time reaching maximum values at 10$^5$-3$\times$10$^5$$\,$years. The gas-phase COM radial profiles show their peak abundances at $\sim$4000$\,$au, {\it which roughly coincides with the position of the CH$_3$OH peak \citep{biz14}}. This is the location where CO starts freezing out onto dust grains in our model and it agrees well with the distance where CO depletion is observed in L1544 \citep[see the drop in C$^{17}$O reported by][]{cas99}. The CO depletion enhances the production of COM precursors on grain surfaces via hydrogenation reactions while, at the same time, the visual extinction is sufficiently high ($A_V$$\sim$7.5-8 mag; Section$\,$\ref{abun}) to prevent the UV photo-dissociation of the chemically desorbed COM precursors by the external interstellar radiation field. UV photo-desorption is not efficient at this position in the core and, therefore, the release of COMs into the gas phase is dominated by chemical reactive desorption. Toward the center of the core, the abundances of COMs drop to undetectable levels ($\leq$10$^{-14}$) as a result of the severe freeze out.  

To compare the model results with our observations, one needs to sample all COM material along the line-of-sight toward the continuum and the {\it CH$_3$OH} peaks. {\it The amount of COM material sampled in the direction of the {\it CH$_3$OH} peak will be larger than toward the continuum peak.} While toward the core's center COMs are found within a shell $R_{i-1} - R_i$ (with $R$ the radius from the center and $i$ the position in the grid with $i$=1 at the outermost shell), toward the {\it CH$_3$OH} peak COMs are sampled along a circle chord. For the core's center the COM column density, $N(\rm{COM})$, is given by:

\begin{equation}
\label{cc}
\begin{array}{l}
 N(\rm{COM})=2\times\sum_{\rm i=2}^{\rm n} \left[\frac{n(H)_i * \chi_i + n(H)_{i-1} * \chi_{i-1}}{2} \right] \times (R_{i-1} - R_i)
\label{Ncont}
\end{array}
\end{equation}

\noindent
with $n$(H)$_i$ the gas density at radial point $i$, $\chi_i$ the COM abundance, $(R_{i-1} - R_i)$ the shell width, and $n$ the number of shells in the model ($n$=129). Toward the {\it CH$_3$OH peak}, $N(\rm{COM})$ is calculated as:

\begin{equation}
\label{cc}
\begin{array}{l}
N(\rm{COM})=2 \times \sum_{\rm i=2}^{\rm n_{peak}} \left[\sqrt{R_{i-1}^2 - R_{peak}^2} - \sqrt{R_{i}^2 - R_{peak}^2} \right] {_\times} \\
\\ 
\ \ \ \ \ \ \ \ \ \ \ \ \ \ \ \ \ \ \ \ \ \ \ {_\times} \left[\frac{n(H)_i * \chi_i + n(H)_{i-1} * \chi_{i-1}}{2} \right]
 \label{Npeak}
\end{array}
\end{equation}

\noindent
where R$_{peak}$ is the radial distance of the {\it CH$_3$OH peak} (4000$\,$au), n$_{peak}$ is the radial point closest to this peak, and R$_i$ and R$_{i-1}$ are the radii at positions $i$ and $i-1$. $N(\rm{COM})$ are averaged over the beam of the IRAM 30$\,$m telescope ($\sim$26$"$), and the COM abundances are finally calculated by dividing the average $N(\rm{COM})$ by the $N(H_2)$ obtained following the same method (4.0$\times$10$^{22}$$\,$cm$^{-2}$ for the continuum peak and 1.0$\times$10$^{22}$$\,$cm$^{-2}$ for the {\it CH$_3$OH} peak). Note that these values are similar to those used in Section$\,$\ref{abun}. The best match with the observations is obtained at 10$^5$$\,$years.

\begin{figure}
\begin{center}
\includegraphics[angle=270,width=0.7\textwidth]{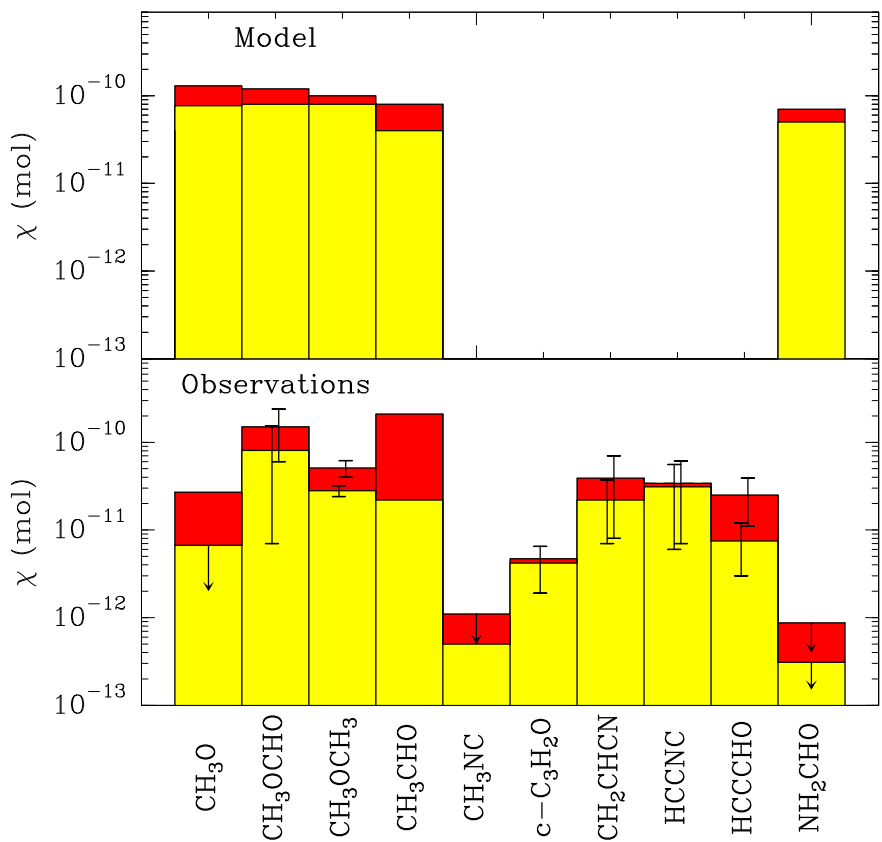}
\caption{Comparison of the observed and modelled abundance of COMs toward the continuum peak (in yellow) and the {\it CH$_3$OH peak} (in red; see Section$\,$\ref{abun}). The abundances shown for c-C$_3$H$_2$O, HCCNC and HCCCHO are average values. Error bars correspond to 1$\sigma$ uncertainties for CH$_3$OCHO, CH$_3$OCH$_3$ and CH$_2$CHCN, while they refer to the lower/upper abundance values estimated for c-C$_3$H$_2$O, HCCNC and HCCCHO (Table$\,$\ref{tab3}). Arrows indicate upper limit abundances.}
\label{fig3}
\end{center}
\end{figure}

Table$\,$\ref{tab3} and Figure$\,$\ref{fig3} (upper panel) report the predicted abundances of CH$_3$O, CH$_3$OCHO, CH$_3$OCH$_3$, CH$_3$CHO and NH$_2$CHO. The rest of O-bearing species (c-C$_3$H$_2$O, HCCCHO, CH$_3$NCO and glycine) are currently not included in the chemical network and their predictions are thus not reported. The modelled COM abundances reproduce the enhancements of CH$_3$OCH$_3$ and CH$_3$OCHO observed toward the {\it CH$_3$OH peak} within factors of 3 (Table$\,$\ref{tab3} and Figure$\,$\ref{fig3}). For CH$_3$CHO, however, the model predicts a smaller enhancement than observed, although the modelled and observed CH$_3$CHO abundances agree within factors 2-3 for both positions. Large discrepancies are found for CH$_3$O and NH$_2$CHO (by factors $\sim$5-15 and $\sim$100-200, respectively; Table$\,$\ref{tab3}), which suggests that additional mechanisms are required to supress the production of these COMs in our model.

In summary, we report new detections of COMs in L1544. Species such as CH$_3$O, CH$_3$CHO and CH$_3$OCH$_3$ are enhanced by factors $\sim$2-10 toward the {\it CH$_3$OH peak} with respect to the core's center. Other COMs such as CH$_3$OCHO may also be enhanced with increasing distance within the core (by factors $\leq$3), although their abundance uncertainties are large. Despite the discrepancies found between the observed COM abundances and those predicted in our model, O-bearing COMs are predicted to present an abundance peak at 4000$\,$au, which agrees well with the position of the {\it CH$_3$OH peak} and with the radial distance at which CO depletion is observed. All this shows that high-sensitivity observations of COMs are strongly needed to put stringent constraints on chemical models and to step forward in our understanding of COM chemistry in the ISM.

\acknowledgments

Observations carried out under projects 129-14 and 001-16. IRAM is supported by INSU/CNRS (France), MPG (Germany) and IGN (Spain). We acknowledge the staff at the IRAM 30m telescope for the support provided during the observations. We also thank an anonymous referee for his/her careful reading of the manuscript and H. S. P. M\"uller for his help in the line identification. This research has received funding from the People Programme (Marie Curie Actions) of the EU's Seventh Framework Programme (FP7/2007-2013) under REA grant agreement PIIF-GA-2011-301538, and from the STFC through an Ernest Rutherford Fellowship (proposal ST/L004801/1). AV and PC acknowledge support from the European Research Council (ERC; Project PALs 320620). LT acknowledges partial support from the Italian Ministero dell\'\,Istruzione, Universit\`a e Ricerca through the grant Progetti Premiali 2012 -- iALMA (CUP C52I13000140001) and from Gothenburg Centre of Advanced Studies in Science and Technology through the program {\it Origins of habitable planets}.


\clearpage

\begin{deluxetable}{lccccccccccccc}
\tabletypesize{\scriptsize}
\rotate
\tablecaption{COM transitions covered in our observations and their derived line parameters \label{tab2}}
\tablewidth{0pt}
\tablehead{
\colhead{Species} & Line & \colhead{Frequency} & \colhead{$E_{\rm u}$} & \colhead{$g_{\rm u}$} & \colhead{$A_{\rm ul}$} & \multicolumn{4}{c}{{\bf (0$"$,0$"$)}} & \multicolumn{4}{c}{{\bf CH$_3$OH peak}} \\ \cline{7-14}
& & & & & & \colhead{Area\tablenotemark{b}} & \colhead{V$_{LSR}$} & \colhead{$\Delta v$} & \colhead{T$_{mb}$\tablenotemark{c}} & \colhead{Area\tablenotemark{b}} & \colhead{V$_{LSR}$} & \colhead{$\Delta v$} & \colhead{T$_{mb}$\tablenotemark{c}} \\
& & (MHz) & (K) & & (s$^{-1}$) & \colhead{(mK km s$^{-1}$)} & \colhead{(km s$^{-1}$)} & \colhead{(km s$^{-1}$)} & \colhead{(mK)} & \colhead{(mK km s$^{-1}$)} & \colhead{(km s$^{-1}$)} & \colhead{(km s$^{-1}$)} & \colhead{(mK)}}
\startdata

CH$_3$O\tablenotemark{a} & F=1$\rightarrow$0, $\Lambda$=-1 & 82455.98 & 4.0 & 3 & 6.5$\times$10$^{-6}$ & $\leq$3.7 & $\ldots$ & $\ldots$ & $\leq$16.2 & 3.5(0.9) & 7.06(0.03) & 0.3(0.1) & 10.8(2.4) \\
	& F=2$\rightarrow$1, $\Lambda$=-1 & 82458.25 & 4.0 & 5 & 9.8$\times$10$^{-6}$ & $\leq$3.7 & $\ldots$ & $\ldots$ & $\leq$15.9 & 9.2(1.1) & 7.21(0.02) & 0.40(0.05) & 21.8(3.0) \\
	& F=2$\rightarrow$1, $\Lambda$=+1 & 82471.82 & 4.0 & 5 & 9.8$\times$10$^{-6}$ & $\leq$3.9 & $\ldots$ & $\ldots$ & $\leq$17.1 & 7.8(1.0) & 7.22(0.02) & 0.30(0.05) & 24.3(2.9) \\
	& F=1$\rightarrow$0, $\Lambda$=+1 & 82524.18 & 4.0 & 3 & 6.5$\times$10$^{-6}$ & $\leq$3.5 & $\ldots$ & $\ldots$ & $\leq$15.0 & 2.6(0.9) & 7.14(0.06) & 0.27(0.08) & 8.9(3.0) \\ \hline
CH$_3$OCHO & 9$_{1,9}$$\rightarrow$8$_{1,8}$ E & 100078.608 & 25.0 & 38 & 1.4$\times$10$^{-5}$ & 1.8(0.8) & 7.23(0.10) & 0.42(0.2) & 4.1(2.0) & $\leq$1.7 & $\ldots$   & $\ldots$   & $\leq$8.1 \\
	   & 9$_{1,9}$$\rightarrow$8$_{1,8}$ A & 100080.542 & 25.0 & 38 & 1.4$\times$10$^{-5}$ & 2.0(1.0) & 7.22(0.19) & 0.44(0.15) & 4.4(1.7) & 3.3(0.8)  & 7.16(0.03) & 0.30(0.09) & 10.4(2.6) \\
	   & 8$_{3,5}$$\rightarrow$7$_{3,4}$ E & 100294.604 & 27.4 & 34 & 1.3$\times$10$^{-5}$ & $\leq$1.3 & $\ldots$ & $\ldots$ & $\leq$5.4 & 3.5(0.9)  & 7.20(0.04) & 0.34(0.10) &  9.9(2.5) \\
	   & 8$_{3,5}$$\rightarrow$7$_{3,4}$ A & 100308.179 & 27.4 & 34 & 1.3$\times$10$^{-5}$ & $\leq$1.7 & $\ldots$ & $\ldots$ & $\leq$7.2 & $\leq$1.4 & $\ldots$   & $\ldots$   & $\leq$6.9 \\
	   & 8$_{1,7}$$\rightarrow$7$_{1,6}$ E & 100482.241 & 22.8 & 34 & 1.4$\times$10$^{-5}$ & $\leq$4.1 & $\ldots$ & $\ldots$ & $\leq$5.7 & 6.5(0.9)  & 7.17(0.02) & 0.31(0.06) & 19.6(2.7) \\
	   & 8$_{1,7}$$\rightarrow$7$_{1,6}$ A & 100490.682 & 22.8 & 34 & 1.4$\times$10$^{-5}$ & 3.5(0.7) & 7.21(0.04) & 0.37(0.07) & 8.9(2.2) & 3.0(0.7)  & 7.16(0.03) & 0.23(0.06) & 12.3(2.6) \\
	   & 9$_{0,9}$$\rightarrow$8$_{0,8}$ E & 100681.545 & 24.9 & 38 & 1.5$\times$10$^{-5}$ & 1.6(0.9) & 7.16(0.11) & 0.4(0.2) & 4.1(2.8) & $\leq$1.6 & $\ldots$   & $\ldots$   & $\leq$7.5 \\
	   & 9$_{0,9}$$\rightarrow$8$_{0,8}$ A & 100683.368 & 24.9 & 38 & 1.5$\times$10$^{-5}$ & 4.1(1.1) & 7.12(0.07) & 0.54(0.14) & 7.2(2.7) & 3.3(0.7)  & 7.11(0.03) & 0.24(0.05) & 13.1(2.6) \\ \hline
CH$_3$OCH$_3$ & 4$_{1,4}$$\rightarrow$3$_{0,3}$ EA\tablenotemark{d} & 99324.357 & 10.2 & 36 & 2.2$\times$10$^{-5}$ & 3.4(0.8) & 7.11(0.05) & 0.38(0.10) & 8.4(2.4) & 6.2(1.2) & 7.11(0.03) & 0.36 (0.08) & 16.4(3.5) \\
	      & 4$_{1,4}$$\rightarrow$3$_{0,3}$ AE\tablenotemark{d} & 99324.359 & 10.2 & 54 & 3.3$\times$10$^{-5}$ & $\ldots$ & $\ldots$ & $\ldots$ & $\ldots$ & $\ldots$ & $\ldots$ & $\ldots$ & $\ldots$  \\
	      & 4$_{1,4}$$\rightarrow$3$_{0,3}$ EE & 99325.208 & 10.2 & 144 & 8.9$\times$10$^{-5}$ & 4.0(0.8) & 7.15(0.03) & 0.32(0.07) & 11.55(2.2) & 9.6(1.2) & 7.12(0.02) & 0.39(0.06) & 23.5(3.1) \\
	      & 4$_{1,4}$$\rightarrow$3$_{0,3}$ AA & 99326.058 & 10.2 &  90 & 5.5$\times$10$^{-5}$ & 2.0(0.9) & 7.22(0.09) & 0.3(0.2) & 5.4(2.5) & 3.7(1.1) & 7.12(0.05) & 0.29(0.09) & 12.0(3.7) \\
	      & 6$_{2,5}$$\rightarrow$6$_{1,6}$ EA\tablenotemark{d} & 100460.412 & 24.7 & 52 & 1.8$\times$10$^{-5}$ & $\leq$1.2 & $\ldots$ & $\ldots$ & $\leq$5.8 & $\leq$1.8 & $\ldots$ & $\ldots$ & $\leq$7.8 \\ 
	      & 6$_{2,5}$$\rightarrow$6$_{1,6}$ AE\tablenotemark{d} & 100460.437 & 24.7 & 78 & 2.6$\times$10$^{-5}$ & $\ldots$ & $\ldots$ & $\ldots$ & $\ldots$  & $\ldots$ & $\ldots$ & $\ldots$ & $\ldots$  \\
	      & 6$_{2,5}$$\rightarrow$6$_{1,6}$ EE & 100463.066 & 24.7 & 208 & 7.0$\times$10$^{-5}$ & $\leq$1.2 & $\ldots$ & $\ldots$ & $\leq$5.7 & 1.4(0.8) & 7.16(0.07) & 0.26(0.16) & 5.2(2.6) \\
	      & 6$_{2,5}$$\rightarrow$6$_{1,6}$ AA & 100465.708 & 24.7 & 130 & 4.4$\times$10$^{-5}$ & $\leq$1.3 & $\ldots$ & $\ldots$ & $\leq$6.3 & $\leq$1.8 & $\ldots$ & $\ldots$ & $\leq$7.8 \\ \hline

CH$_3$CHO     & 2$_{1,2}$$\rightarrow$1$_{0,1}$ E  & 83584.260  & 5.0  & 10  & 2.2$\times$10$^{-6}$ & 3.8(1.5) & 7.12(0.04) & 0.32(0.10) & 10.9(2.6) & 12.8(1.1) & 7.18(0.01) & 0.31(0.04) & 39.0(3.1) \\
	      & 2$_{1,2}$$\rightarrow$1$_{0,1}$ A  & 84219.764  & 5.0  & 10  & 2.4$\times$10$^{-6}$ & 5.5(0.8) & 7.21(0.02) & 0.30(0.05) & 17.4(2.5) & 9.6(0.9)  & 7.12(0.01) & 0.26(0.02) & 34.4(3.0) \\
	      & 2$_{1,1}$$\rightarrow$1$_{0,1}$ E  & 87109.504  & 5.2  & 10  & 1.3$\times$10$^{-7}$ & $\leq$1.4 & $\ldots$ & $\ldots$ & $\leq$6.5 & $\leq$1.7 & $\ldots$ & $\ldots$ & $\leq$7.7 \\
	      & 2$_{2,0}$$\rightarrow$3$_{1,3}$ A  & 87146.655  & 11.8 & 10  & 2.9$\times$10$^{-7}$ & $\leq$1.3 & $\ldots$ & $\ldots$ & $\leq$6.2 & $\leq$1.5 & $\ldots$ & $\ldots$ & $\leq$6.6 \\
	      & 2$_{2,0}$$\rightarrow$3$_{1,3}$ E  & 87204.278  & 11.9 & 10  & 1.1$\times$10$^{-7}$ & $\leq$1.6 & $\ldots$ & $\ldots$ & $\leq$5.9 & $\leq$1.7 & $\ldots$ & $\ldots$ & $\leq$7.4 \\
	      & 6$_{3,4}$$\rightarrow$7$_{2,5}$ A  & 99141.294  & 24.4 & 10  & 7.1$\times$10$^{-7}$ & $\leq$1.4 & $\ldots$ & $\ldots$ & $\leq$6.7 & 1.8(0.8)  & 6.95(0.03) & 0.2(0.2) & 8.1(3.0) \\
	      & 6$_{3,4}$$\rightarrow$7$_{2,6}$ E  & 99490.149  & 24.3 & 10  & 6.1$\times$10$^{-7}$ & $\leq$1.4 & $\ldots$ & $\ldots$ & $\leq$6.7 & 2.9(0.9)  & 7.33(0.07) & 0.4(0.1) & 6.3(2.3) \\ \hline
CH$_3$NC      & 5$_0$$\rightarrow$4$_0$            & 100526.541 & 14.5 & 44  & 8.1$\times$10$^{-5}$ & 4.0(1.0)  & 7.15(0.09)   & 0.66(0.18)   & 5.7(2.3)  & $\leq$1.8 & $\ldots$ & $\ldots$ & $\leq$7.5 \\ 	
              & 5$_1$$\rightarrow$4$_1$            & 100524.249 & 21.5 & 44  & 7.8$\times$10$^{-5}$ & 2.6(0.7)  & 7.38(0.05)   & 0.35(0.11)   & 7.0(2.2)  & $\leq$1.8 & $\ldots$ & $\ldots$ & $\leq$7.5 \\ 
              & 5$_2$$\rightarrow$4$_2$            & 100517.433 & 42.7 & 44  & 6.8$\times$10$^{-5}$ & $\leq$1.7 & $\ldots$     & $\ldots$     & $\leq$6.9  & $\leq$1.8 & $\ldots$ & $\ldots$ & $\leq$7.5 \\ \hline
c-C$_3$H$_2$O & 6$_{1,6}$$\rightarrow$5$_{1,5}$\tablenotemark{e}    & 79483.520  & 14.6 & 39  & 5.1$\times$10$^{-5}$ & $\ldots$ & $\ldots$ & $\ldots$ & $\ldots$ & 7.7(1.4) & 7.09(0.03) & 0.38(0.08) & 18.8(3.0) \\     
	      & 7$_{1,6}$$\rightarrow$6$_{1,5}$    & 103069.925 & 21.1 & 45  & 1.1$\times$10$^{-4}$ & 3.3(0.9)  & 7.18(0.06)   & 0.39(0.09)   & 8.0(2.8)  & 3.7(1.7)  & 7.1(0.1)   & 0.5(0.3) & 7.6(3.4) \\ \hline
CH$_2$CHCN    & 9$_{0,9}$$\rightarrow$8$_{0,8}$    & 84946.000  & 20.4 & 57  & 4.9$\times$10$^{-5}$ & 24.78(1.0)& 7.290(0.007) & 0.397(0.019) & 58.7(2.3) & 6.6(1.1)  & 7.30(0.02) & 0.29(0.06) & 21.1(3.1) \\
	      & 9$_{1,8}$$\rightarrow$8$_{1,7}$    & 87312.812  & 23.1 & 57  & 5.3$\times$10$^{-5}$ & 13.2(0.9) & 7.245(0.012) & 0.35(0.03)   & 35.3(2.4) & 5.1(1.1)  & 7.20(0.06) & 0.46(0.09) & 10.4(2.7) \\
	      &11$_{0,11}$$\rightarrow$10$_{0,10}$ & 103575.395 & 29.9 & 69  & 9.0$\times$10$^{-5}$ & 7.8(0.8)  & 7.30(0.02)   & 0.42(0.05)   & 17.3(2.3) & $\leq$1.8 & $\ldots$ & $\ldots$ & $\leq$8.7 \\ \hline
HCCNC	      & 10$\rightarrow$9\tablenotemark{f}  & 99354.250  & 26.2 & 63  & 4.6$\times$10$^{-5}$ & 29.2(1.0) & 7.227(0.005) & 0.438(0.013) & 62.7(1.6) & 7.1(1.0)  & 7.17(0.03) & 0.40(0.05) & 16.5(2.2) \\ \hline
HCCCHO 	      & 9$_{0,9}$$\rightarrow$8$_{0,8}$    & 83775.842  & 20.1 & 19  & 1.8$\times$10$^{-5}$ & 5.5(1.2)  & 7.08(0.06)   & 0.49(0.11)   & 10.6(2.8) & 3.2(1.2)  & 7.04(0.06) & 0.28(0.12) & 10.8(3.5) \\ \hline
NH$_2$CHO & 4$_{0,4}$$\rightarrow$3$_{0,3}$\tablenotemark{g} & 84542.400 & 10.2 & 11 & 4.1$\times$10$^{-5}$ & $\leq$1.8 & $\ldots$ & $\ldots$ & $\leq$8.4 & $\leq$2.3 & $\ldots$ & $\ldots$ & $\leq$9.6 \\
	  & 4$_{1,3}$$\rightarrow$3$_{1,2}$\tablenotemark{g} & 87848.915 & 13.5 & 11 & 4.3$\times$10$^{-5}$ & $\leq$1.3 & $\ldots$ & $\ldots$ & $\leq$6.0 & $\leq$1.8 & $\ldots$ & $\ldots$ & $\leq$7.2 \\ \hline
CH$_3$NCO & 10$_{1,10}$$\rightarrow$9$_{1,9}$ & 87506.605 & 29.0 & 21 & 3.0$\times$10$^{-5}$ & $\leq$1.1 & $\ldots$ & $\ldots$ & $\leq$5.1 & $\leq$1.2 & $\ldots$ & $\ldots$ & $\leq$5.7 \\
	  & 12$_{-1,12}$$\rightarrow$11$_{-1,11}$ & 103023.61 & 39.4 & 25 & 4.9$\times$10$^{-5}$ & $\leq$1.3 & $\ldots$ & $\ldots$ & $\leq$6.3 & $\leq$1.6 & $\ldots$ & $\ldots$ & $\leq$7.5 \\ \hline
Glycine   & 6$_{5,2}$$\rightarrow$5$_{4,1}$ & 103294.648 & 15.2 & 39 & 1.6$\times$10$^{-6}$ & $\leq$1.7 & $\ldots$ & $\ldots$ & $\leq$8.0 & $\leq$2.1 & $\ldots$ & $\ldots$ & $\leq$9.9 \\
Conf I.   & 6$_{5,1}$$\rightarrow$5$_{4,2}$ & 103297.993 & 15.2 & 39 & 1.6$\times$10$^{-6}$ & $\leq$1.7 & $\ldots$ & $\ldots$ & $\leq$8.0 & $\leq$2.1 & $\ldots$ & $\ldots$ & $\leq$9.9 

\enddata

\tablenotetext{a}{Hyperfine components of the N=1-0, K=0, J=3/2$\rightarrow$1/2 transition of CH$_3$O.}
\tablenotetext{b}{Upper limits calculated as 3$\sigma$$\times\sqrt{\Delta v \times \delta v}$, with $\sigma$ the rms noise level, $\Delta$v the linewidth and $\delta$v the velocity resolution of the spectrum.}
\tablenotetext{c}{The rms noise level, $\sigma$, is given in parenthesis. Upper limits to the peak intensities refer to the 3$\sigma$ noise level}
\tablenotetext{d}{The AE and EA transitions overlap. We only show the gaussian fit for one of the lines. In the rotational diagram of this species, the individual areas of the AE and EA transitions are calculated by weighting the blended area by the degeneracy g$_u$$\times$g$_i$ of every transition.}
\tablenotetext{e}{Transition not observed toward the continuum peak.}
\tablenotetext{f}{Hyperfine structure not resolved.}
\tablenotetext{g}{Hyperfine transitions blended. Spectroscopic information provided only for the brightest hyperfine component F=5-4.}

\end{deluxetable}

\begin{deluxetable}{lcccc|cccc}
\tabletypesize{\scriptsize}
\rotate
\tablecaption{Excitation temperatures, column densities, and measured and modelled abundances of COMs in L1544 \label{tab3}}
\tablewidth{0pt}
\tablehead{
\colhead{Species} & \multicolumn{4}{c}{{\bf (0$"$,0$"$)}} & \multicolumn{4}{c}{{\bf CH$_3$OH peak}} \\ 
& \colhead{T$_{ex}$ (K)} & \colhead{N$_{obs}$ (cm$^{-2}$)} & \colhead{$\chi_{\rm obs}$\tablenotemark{c}} & \colhead{$\chi_{\rm mod}$\tablenotemark{d}} & \colhead{T$_{ex}$ (K)} & \colhead{N$_{obs}$ (cm$^{-2}$)} & \colhead{$\chi_{\rm obs}$\tablenotemark{c}} & \colhead{$\chi_{\rm mod}$\tablenotemark{d}}}
\startdata

CH$_3$O    	     & 5-10\tablenotemark{b} & $\leq$(2.8-3.6)$\times$10$^{11}$ & $\leq$(5.1-6.7)$\times$10$^{-12}$ & 7.7$\times$10$^{-11}$ & 8.0\tablenotemark{e} & 4.0$\times$10$^{11}$ & 2.7$\times$10$^{-11}$ & 1.3$\times$10$^{-10}$ \\
CH$_3$OCHO           & 5.1$\pm$2.3\tablenotemark{a} & (4.4$\pm$4.0)$\times$10$^{12}$\tablenotemark{a} & (8.1$\pm$7.4)$\times$10$^{-11}$ & 8.0$\times$10$^{-11}$ & 7.9$\pm$3.6 & (2.3$\pm$1.4)$\times$10$^{12}$ & (1.5$\pm$0.9)$\times$10$^{-10}$ & 1.2$\times$10$^{-10}$ \\ 
CH$_3$OCH$_3$        & 5.7$\pm$3.1 & (1.5$\pm$0.2)$\times$10$^{12}$ & (2.8$\pm$0.4)$\times$10$^{-11}$ & 8.0$\times$10$^{-11}$ & 7.6$\pm$3.7 & (7.7$\pm$1.6)$\times$10$^{11}$ & (5.1$\pm$1.1)$\times$10$^{-11}$ & 1.0$\times$10$^{-10}$ \\
CH$_3$CHO     	     & 5.0\tablenotemark{e} & 1.2$\times$10$^{12}$ & 2.2$\times$10$^{-11}$ & 4.0$\times$10$^{-11}$ & 7.8\tablenotemark{e} & 3.2$\times$10$^{12}$ & 2.1$\times$10$^{-10}$ & 8.0$\times$10$^{-11}$ \\
CH$_3$NC             & 22.9\tablenotemark{e} & 2.7$\times$10$^{10}$ & 5.0$\times$10$^{-13}$ & $\ldots$ & 5-10\tablenotemark{b} & $\leq$(0.7-1.6)$\times$10$^{10}$ & $\leq$(0.5-1.1)$\times$10$^{-12}$ & $\ldots$ \\      
c-C$_3$H$_2$O\tablenotemark{f}        & 5-10\tablenotemark{b} & (1.0-3.5)$\times$10$^{11}$ & (1.9-6.5)$\times$10$^{-12}$ & $\ldots$ & 8.0\tablenotemark{e} & 7.0$\times$10$^{10}$ & 4.7$\times$10$^{-12}$ & $\ldots$ \\ 
CH$_2$CHCN           & 5.8$\pm$0.9 & (1.2$\pm$0.8)$\times$10$^{12}$ & (2.2$\pm$1.5)$\times$10$^{-11}$ & $\ldots$ & 5.0$\pm$1.4 & (5.8$\pm$4.7)$\times$10$^{11}$ & (3.9$\pm$3.1)$\times$10$^{-11}$ & $\ldots$ \\
HCCNC & 5-10\tablenotemark{b} & (0.3-3.0)$\times$10$^{12}$ & (0.6-5.6)$\times$10$^{-11}$ & $\ldots$ & 5-10\tablenotemark{b} & (1.0-9.1)$\times$10$^{11}$ & (0.7-6.1)$\times$10$^{-11}$ & $\ldots$ \\	      
HCCCHO & 5-10\tablenotemark{b} & (1.8-6.3)$\times$10$^{11}$ & (0.3-1.2)$\times$10$^{-11}$ & $\ldots$ & 5-10\tablenotemark{b} & (1.6-5.8)$\times$10$^{11}$ & (1.1-3.9)$\times$10$^{-11}$ & $\ldots$ \\	
NH$_2$CHO & 5-10\tablenotemark{b} & $\leq$(1.3-1.7)$\times$10$^{10}$ & $\leq$(2.4-3.1)$\times$10$^{-13}$ & 5.0$\times$10$^{-11}$ & 5-10\tablenotemark{b} & $\leq$(1.0-1.3)$\times$10$^{10}$ & $\leq$(6.7-8.7)$\times$10$^{-13}$ & 7.0$\times$10$^{-11}$ \\
CH$_3$NCO & 5-10\tablenotemark{b} & $\leq$(0.8-6.3)$\times$10$^{11}$ & $\leq$(0.2-1.2)$\times$10$^{-11}$ & $\ldots$ & 5-10\tablenotemark{b} & $\leq$(0.9-6.3)$\times$10$^{11}$ & $\leq$(0.6-4.2)$\times$10$^{-11}$ & $\ldots$ \\
Glycine & 5-10\tablenotemark{b} & $\leq$(3.3-5.8)$\times$10$^{12}$ & $\leq$(0.6-1.1)$\times$10$^{-10}$ & $\ldots$ & 5-10\tablenotemark{b} & $\leq$(4.2-9.5)$\times$10$^{12}$ & $\leq$(2.8-6.3)$\times$10$^{-10}$ & $\ldots$

\enddata

\tablenotetext{a}{Errors correspond to 1$\sigma$ uncertainties in MADCUBAIJ.}
\tablenotetext{b}{T$_{ex}$ range assumed to calculate the total column densities from a single COM transition using MADCUBAIJ.}
\tablenotetext{c}{Molecular abundances calculated by using an H$_2$ column density of 5.4$\times$10$^{22}$$\,$cm$^{-2}$ for the continuum peak (see Section$\,$\ref{abun} for details) and of 1.5$\times$10$^{22}$$\,$cm$^{-2}$ for the position of the {\it CH$_3$OH peak} \citep{spez16}}.
\tablenotetext{d}{Abundances predicted by the model of Vasyunin et al. (in prep).}
\tablenotetext{e}{Fitting solutions could only be obtained by fixing T$_{ex}$ within MADCUBAIJ.}
\tablenotetext{f}{T$_{ex}$ and N$_{obs}$ calculated using Weeds.}

\end{deluxetable}

\end{document}